\documentstyle[epsfig,12pt]{article}
      
\begin{document}

\vspace*{-15mm}

\begin{center}
  {\Large {\bf Topologically Massive Abelian Gauge Theory }} \\[5mm] 
  {\large K. Saygili\footnote{Electronic address: ksaygili@yeditepe.edu.tr}}
  \\[3mm]
  {Department of Mathematics, Yeditepe University,} \\ 
  {Kayisdagi, 34755 Istanbul, Turkey} \\[10mm]

  Abstract: \\

\end{center}

  We discuss three mathematical structures which arise in topologically 
massive abelian gauge theory. First, the euclidean topologically massive 
abelian gauge theory defines a contact structure on a manifold. We briefly 
discuss three solutions and the related contact structures on the flat 
$3$-torus, the AdS space, the $3$-sphere which respectively correspond 
to Bianchi type $I$, $VIII$, $IX$  spaces. We also present solutions 
on Bianchi type $II$, $VI$ and $VII$ spaces. Secondly, we discuss a
family of complex (anti-)self-dual solutions of the euclidean theory 
in cartesian coordinates on ${\mathcal{R}}^{3}$ which are given by 
(anti-)holomorpic functions. The orthogonality relation of contact 
structures which are determined by the real parts of these complex 
solutions separates them into two classes: the self-dual and the 
anti-self-dual solutions. Thirdly, we apply the curl transformation 
to this theory. An arbitrary solution is given by a vector tangent 
to a sphere whose radius is determined by the topological mass in 
transform space. Meanwhile a gauge transformation corresponds 
to a vector normal to this sphere. We discuss the quantization 
of topological mass on an example.

\section{Introduction}

   Topologically massive gauge theories are dynamical theories which 
are specific to three dimensions \cite{DJTS1, DJTS2}, \cite{DJTS3}. 
They are qualitatively different from the Yang-Mills type gauge theories 
beside their mathematical elegance and consistency. They provide an 
alternative way to introduce mass term with no spontaneous symmetry breaking. 
The most distinctive feature of the topologically massive gauge theories is 
the existence of a natural scale of length which is determined by the inverse 
of the topological mass \cite{D, J}, \cite{ANS, S, S1}. This leads to purely
geometric discussions of these theories \cite{S, S1}.
 
   We discuss three mathematical structures in this context. The effect of 
the topological mass is to introduce a natural scale of length into these 
structures. To the knowledge of the author, the connection of these structures 
with the topologically massive gauge theories has been overlooked in the 
literature. Our results further contribute to discussions of these
structures.

  First we discuss the connection of the topologically massive abelian 
gauge theory with contact geometry. A real-valued topologically massive 
abelian gauge potential on a Riemannian (euclidean signature) manifold 
is a Beltrami (Trkalian) field \cite{AK} which defines a contact structure. 
We discuss three solutions and the relevant contact structures arising on 
the flat $3$-torus ${\mathcal{T}}^{3}$, the AdS space (lorentzian) 
${\mathcal{H}}^{3}$, the $3$-sphere ${\mathcal{S}}^{3}$ which respectively 
correspond to Bianchi type $I$, $VIII$, $IX$ spaces. We also present 
solutions on Bianchi type $II$, $VI$ and $VII$ spaces. We realize the 
Darboux contact form as a topologically massive gauge potential in 
euclidean (type $II$ and $VII$) and lorentzian (type $VI$) signatures. 
The Bianchi type $II$, $VI$, $VII$, $VIII$ and $IX$ spaces are homogeneous 
contact manifolds \cite{P}, \cite{RS, MC}. We further present complex-valued 
or lorentzian solutions which do not lead to contact structures on Bianchi 
type $V$ and specialized forms of Bianchi type $VI$ and $VII$ spaces.

  Secondly, we discuss a family of complex Beltrami (Trkalian) fields 
including the topological mass in cartesian coordinates on ${\mathcal{R}}^{3}$ 
\cite{BT}. The complex (anti-)self-dual solutions of the euclidean
(and also lorentzian) theory are given by (anti-)holomorpic functions. 
The orthogonality relation of contact structures which are determined 
by the real parts of these complex solutions separates them into two 
classes: the self-dual and the anti-self-dual solutions. These two 
classes, in real case, possess opposite helicities.

   Thirdly, we apply the curl transformation \cite{M1}, \cite{Ml1, Ml3} 
to this theory. This is based on decomposing a solution into (complex) 
helical eigen-functions of the curl operator in the fashion of a Fourier 
transform refining the Helmholtz's decomposition. An arbitrary field 
is given by a vector tangent to a sphere whose radius is determined by 
the topological mass in transform space \cite{Ml1}, \cite{MD1}. 
Meanwhile a gauge transformation corresponds to a vector normal to 
this sphere in this space. We also discuss the sense of quantization 
of topological mass on an example.

   The references \cite{MD1, MD3} and \cite{K1, K2} contain nice, elementary
expositions of some of the relevant basic mathematical concepts in different 
contexts.  

\section{Contact Structures and Beltrami Fields}

  A contact form $\alpha$ on a three dimensional manifold ${\mathcal{M}}$ 
is a $1$-form such that $\alpha \wedge d\alpha$ vanishes nowhere: 
$\alpha \wedge d\alpha \neq 0$. A contact structure $\xi$ is a smooth 
tangent plane field on ${\mathcal{M}}$ which is locally given as the kernel 
of a contact $1$-form: $\xi= ker( \alpha )$. Thus a manifold ${\mathcal{M}}$ 
endowed with a globally defined contact structure is called a contact manifold 
\cite{Arnold}, \cite{Blair, OS}, \cite{L, E, HG}. A tangent plane in $\xi$ at 
a point of ${\mathcal{M}}$ is a contact element. It follows from the Frobenius 
integrability theorem that a contact structure is a maximally non-integrable 
tangent plane field on ${\mathcal{M}}$. The unique vector field 
$\overrightarrow{X}$ which satisfies $\alpha(\overrightarrow{X})=1$ and 
${\mathcal{L}}_{\overrightarrow{X}}\alpha=\overrightarrow{X} \rfloor d\alpha 
=0$ is called the characteristic or Reeb vector field associated with the 
contact structure. Further if the Reeb vector is a Killing vector: 
${\mathcal{L}}_{ \overrightarrow{X} } g=0$ for a Riemannian metric $g$ 
on ${\mathcal{M}}$ then this is called a K-contact structure \cite{Blair}. 

   A diffeomorphism $f$ on a contact manifold ${\mathcal{M}}$ is a contact 
transformation (contactomorphism) if it preserves the contact structure. 
More precisely, two contact structures $\xi_{\alpha}= ker( \alpha )$ and 
$\xi_{\beta}= ker( \beta )$ on a manifold ${\mathcal{M}}$ are contactomorphic 
if there exists a contact transformation $f$ so that $f^{*}\beta = \mu\alpha$ 
for some non-vanishing function $\mu$ on ${\mathcal{M}}$ or equivalently 
$f_{*}\xi_{\alpha}=\xi_{\beta}$ \cite{OS}. 

  Darboux theorem asserts that there exists local coordinates about any 
point on a contact manifold ${\mathcal{M}}$ in which the contact form can 
be expressed as $\alpha=dz+xdy$ \cite{Arnold}. In other words all contact 
structures are locally contactomorphic. Thus they locally look alike in 
Darboux coordinates \cite{EG1}. However these are implicitly global objects 
\cite{G}. 

  The eigen-vectors of the curl operator 

\begin{eqnarray}\label{Beltramivector}
\overrightarrow{\nabla} \times \overrightarrow{\alpha} = 
\lambda \overrightarrow{\alpha} ,
\end{eqnarray}

\noindent on ${\mathcal{R}}^{3}$ are called Beltrami vectors \cite{AK}. 
The eigen-value $\lambda$ is a function on ${\mathcal{R}}^{3}$. If this 
is a constant then the eigen-vector is called Trkalian \cite{BT}. We
refer the reader to \cite{W} for a derivation from a variational 
principle of this equation for constant eigen-value. We can 
associate a $1$-form $\alpha=\alpha_{i}dx^{i}$ with the vector 
$\overrightarrow{\alpha}=\alpha^{i}\partial_{i}$ if we \textit{adapt} 
the metric $g_{ij}=diag(1, \, 1, \, 1)$ \cite{CH}. The metric induces 
the usual correspondence between vectors and $1$-forms. Then we can 
write the equation (\ref{Beltramivector}) using differential forms as

\begin{eqnarray}\label{Beltramiform}
*d\alpha - \lambda \alpha=0 , 
\end{eqnarray}

\noindent ($\overrightarrow{\nabla} \times \leftrightarrow *d$). 
If $\alpha$ is Trkalian then $d*\alpha=0$, ($\overrightarrow{\nabla} 
\cdot \leftrightarrow d*$). Thus a Beltrami $1$-form $\alpha$ defines 
a contact structure since $\alpha \wedge d\alpha=\lambda \alpha 
\wedge *\alpha \neq 0$ \cite{EG1}. This expression defines a volume 
form on ${\mathcal{R}}^{3}$. The  contact structure is positively 
(negatively) oriented if this orientation agrees (disagrees) with 
the orientation on ${\mathcal{R}}^{3}$, depending on the sign of 
$\lambda$.

   The helicity of the $1$-form $\alpha$ is defined as

\begin{eqnarray}\label{helicity}
H (\alpha)=\int_{D} \alpha \wedge d\alpha ,
\end{eqnarray}

\noindent where $D$ is a domain in ${\mathcal{R}}^{3}$ \cite{AK}. The 
helicity density ${\mathcal{H}} (\alpha)=\alpha \wedge d \alpha$ is a 
local measure of \textit{twisting} of the smooth contact structure 
defined by $\alpha$ so as to be maximally non-integrable \cite{MD1}. 
The contact field has positive (negative) helicity if 
$\lambda>0$ ($\lambda<0$). The helicity (\ref{helicity}) 
corresponds to an abelian Chern-Simons term in gauge theory 
\cite{J1, J2}.

   It has been frequently observed in the literature that a curl 
eigen-vector is dual to a contact form \cite{G, EG2, EG3, EG4} (and the 
references therein), (see also \cite{MD1, MD3} and \cite{K1, K2}). 
A more precise correspondence follows by adapting a metric to a 
contact form as introduced in \cite{CH}. This is also related to 
contact metric structures \cite{Blair}. A Riemannian metric is said 
to be adapted to the contact form $\alpha$ if $\alpha$ is of unit 
\textit{length} and it satisfies the equation (\ref{Beltramiform}) 
for $\lambda=2$ \cite{CH}. We shall use a slightly more general 
definition by unrestricting the length of $\alpha$ and $\lambda$.

   A homogeneous manifold with an invariant contact structure
is called a homogeneous contact manifold. More precisely, if there 
exist a metric associated with the contact $1$-form $\alpha$ and 
a group of diffeomorphisms acting transitively as a group of 
isometries which leave $\alpha$ invariant then we have a 
homogeneous contact manifold \cite{P}. We refer the reader 
to \cite{CH} for the existence of adapted metrics on any 
Riemannian contact manifold. 

  The self-duality equation in three dimensions

\begin{eqnarray}\label{selfduality}
(*d - \nu) A=*F-\nu A=0 ,
\end{eqnarray}

\noindent \cite{TPN, Deser1}, is the generalization of the eigen-form
equation (\ref{Beltramiform}) of the $*d$ operator to a curved manifold 
${\mathcal{M}}$ with an adapted metric $g_{\mu\nu}$ and a gauge potential 
$A=\alpha$ on it, for a constant eigen-value $\nu=\lambda$. The 
field equation of the topologically massive abelian gauge theory 

\begin{eqnarray}\label{field equation}
(*d-\nu) *F = *d(*d-\nu) A=0 ,
\end{eqnarray}

\noindent is given by applying the operator $*d$ on the self-duality 
equation (\ref{selfduality}), \cite{S}. Thus a real-valued self-dual 
solution of the topologically massive abelian gauge theory on a 
Riemannian (euclidean signature) manifold ${\mathcal{M}}$ is given 
by a Beltrami (Trkalian) gauge potential. Accordingly this gauge 
potential defines a contact structure on ${\mathcal{M}}$.
We remark that $\alpha\wedge*\alpha$ is positive definite on a 
Riemannian manifold ${\mathcal{M}}$ if $\alpha$ is real-valued. 
But this is not true if $\alpha$ is complex-valued or ${\mathcal{M}}$
is of lorentzian signature. We shall call: $(*d + \nu) A=*F+\nu A=0$ 
the anti-self-duality equation in anticipation with (\ref{selfduality}).
The examples below consist of solutions on Bianchi type spaces. These 
are homogeneous spaces.

  Furthermore, the contact structure defined by the $1$-form $*F$ in 
(\ref{selfduality}) coincides with that of $A$. After all, a contact 
structure is defined up to a multiple ($\nu$) of the contact $1$-from 
$\alpha=A$ ($\beta=\nu A = *dA=*F$). Note that the field equation 
(\ref{field equation}) is simply the self-duality equation written
for the contact $1$-form $*F$ since the equations (\ref{selfduality}) 
and (\ref{field equation}) are symmetric under the interchange 
$\nu A \leftrightarrow *F$, \cite{S}.  The helicity densities 
of the potential $A$ and the dual-field $*F$ are related as 
${\mathcal{H}} (*F)={\nu}^{2} {\mathcal{H}} (A)$. The helicity
density is a gauge-dependent quantity. The self-dual model which is 
separately introduced in \cite{TPN} is related to the topologically 
massive abelian theory by a Legendre transformation \cite{Deser1}.

 The contact structure defined by the gauge potential $A$ (or dual-field 
$*F$) is locally contactomorphic to the contact structure defined by the 
gauge Darboux form $A^{\prime}=f^{*}A=dz+xdy$. In other words there exists 
local coordinates in which the topologically massive contact gauge potential 
$1$-form is given by $A^{\prime}$. Then the self-duality equation in these 
coordinates 

\begin{eqnarray}
(*_{g^{\prime}}d - \nu) A^{\prime}=0 ,
\end{eqnarray}

\noindent is also satisfied adapting the pull-back metric $g^{\prime}=f^{*}g$ 
since $f^{*}*_{g}dA=*_{g^{\prime}}f^{*}dA=*_{g^{\prime}}df^{*}A$.
Here $f$ is an orientation preserving contactomorphism.

\subsection{The $3$-sphere}

   A self-dual solution of the euclidean topologically massive 
abelian gauge theory on the $3$-sphere ${\mathcal{S}}^{3}$ is 
given in \cite{ANS}, \cite{S}. The $3$-sphere which is locally 
given as ${\mathcal{S}}^{1} \times {\mathcal{S}}^{2}$ is a Bianchi 
type $IX$ space \cite{ANS}. The effect of the topological mass is 
to introduce a natural scale of length $r=2/\nu$ where $\nu=ng^{2}$ 
\cite{S}. The contact gauge potential is given by

\begin{eqnarray}\label{euclidgaugepotential}
A=-\frac{1}{2}\frac{\nu}{g} \, \omega^{3} =-\frac{1}{2}\frac{\nu}{g} 
\, \left[ d\psi + \cos ( \nu\theta ) d\phi \right] .
\end{eqnarray}
 
\noindent Here

\begin{eqnarray}
& & \omega^{1} = -\sin(\nu\psi)d\theta+\cos(\nu\psi)\sin(\nu\theta)d\phi ,
\hspace*{10mm} \eta_{ab}=diag(1, \, 1, \, 1) , \nonumber \\
& & \omega^{2} = \cos(\nu\psi)d\theta+\sin(\nu\psi)\sin(\nu\theta)d\phi , \\
& & \omega^{3} = d\psi+\cos(\nu\theta)d\phi , \nonumber 
\end{eqnarray}

\noindent are the \textit{modified} left-invariant basis $1$-forms of 
$SU(2)$ which is parameterized in terms of the Eulerian (half) arclengths 
$\theta$, $\phi$ and $\psi$ corresponding to the Euler angles 
$\tilde{\theta}=\nu\theta$, $\tilde{\phi}=\nu\phi$ and 
$\tilde{\psi}=\nu\psi$ on the $3$-sphere ${\mathcal{S}}^{3}$ 
of radius $r$ \cite{S}. The metric is given by

\begin{eqnarray} \label{metric}
& & ds^{2}=\eta_{ab}\omega^{a}\omega^{b} .
\end{eqnarray}

\noindent The Maurer-Cartan equation: $d\omega^{a}
=-\frac{1}{2}C_{b \,\, c}^{\,\, a} \omega^{b} \wedge \omega^{c}$ 
for the basis $1$-forms yields the non-vanishing structure constants:
$C_{2 \,\, 3}^{\,\, 1}=-\nu$, $C_{3 \,\, 1}^{\,\, 2}=-\nu$, 
$C_{1 \,\, 2}^{\,\, 3}=-\nu$ of the Bianchi type $IX$ group which 
are modified by the topological mass $\nu$.  The Maurer-Cartan equation 
and Hodge duality relations for the basis $1$-forms immediately lead to 
the result that $\alpha=A$ satisfies the self-duality equation.

  The $1$-form $\alpha=A$ defines the standard contact structure on 
${\mathcal{S}}^{3}$ \cite{OS}, \cite{L, E, HG, EG4}. The contact plane field 
$\xi=ker(A)=Span\{ \overrightarrow{e}_{1}, \, \overrightarrow{e}_{2} \}$ 
is orthogonal to the trajectory of the vector $\overrightarrow{e}_{3}$ which 
is a Hopf fiber \cite{G}. Here 
$\{ \overrightarrow{e}_{1}, \, \overrightarrow{e}_{2}, \, 
\overrightarrow{e}_{3} \}$ is the frame dual to the co-frame 
$\{ \omega^{1}, \, \omega^{2}, \, \omega^{3} \}$. Further the Reeb 
vector is given by $\overrightarrow{e}_{3}$  \cite{EG1}, which is also 
a Killing vector for the metric. Thus the Hopf contact gauge potential $A$ 
defines a K-contact structure on ${\mathcal{S}}^{3}$. Note that $\omega^{1}$ 
and $\omega^{2}$ are also self-dual Beltrami (Trkalian) $1$-forms \cite{ANS}. 
These also define contact structures on ${\mathcal{S}}^{3}$ \cite{T}.

  The scaling of the unmodified basis $1$-forms by the radius of the 
$3$-sphere is also noted in \cite{N}, \cite{Gu}.

\subsection{The Flat $3$-Torus}

  The next example of a topologically massive [Beltrami (Trkalian)] gauge
potential is the \textit{abc-flow} on the flat $3$-torus ${\mathcal{T}}^{3}$ 
\cite{AK}, \cite{EG1}. Consider the \textit{abc-potential}  

\begin{eqnarray}\label{ABCcontactgauge}
& & A=\frac{\nu}{g} \left( Udx+Vdy+Wdz \right) , \nonumber \\
\\
& & U=a\sin (\nu z) + c\cos (\nu y) , \nonumber \\
& & V=b\sin (\nu x) + a\cos (\nu z) , \nonumber \\
& & W=c\sin (\nu y) + b\cos (\nu x) , \nonumber 
\end{eqnarray}

\noindent with the topological mass $\nu$ on the flat $3$-torus 
${\mathcal{T}}^{3}$. Here $a$, $b$ and $c$ are arbitrary constants. 
The flat $3$-torus  

\begin{eqnarray}
{\mathcal{T}}^{3}= \{ (x, \, y, \, z) \in R^{3}; 
mod \, 2\pi/\nu \} ,
\end{eqnarray}

\noindent is given by identifying the opposite faces of a cube of 
edge length $2\pi/\nu$ in ${\mathcal{R}}^{3}$. This can be embedded 
in a five-dimensional sphere ${\mathcal{S}}^{5}$ in ${\mathcal{R}}^{6}$ 
\cite{Blair}. The metric inherited from the  natural flat metric on 
${\mathcal{R}}^{6}$ coincides with the flat metric on ${\mathcal{R}}^{3}$ 
in which the cube resides. Thus this (more precisely the covering space 
${\mathcal{R}}^{3}$) is a Bianchi type $I$ space

\begin{eqnarray}
& & \omega^{1} = dx , \hspace*{10mm} \eta_{ab}=diag(1, \, 1, \, 1) , 
\nonumber \\
& & \omega^{2} = dy , \\
& & \omega^{3} = dz. \nonumber 
\end{eqnarray}

\noindent The effect of the topological mass is again to introduce a 
natural scale of length $r=1/\nu$ where $\nu=ng^{2}$. The parameters 
$x$, $y$ and $z$ are the arclengths corresponding to the angles 
$\tilde{x}=\nu x$, $\tilde{y}=\nu y$, $\tilde{z}=\nu z$ on 
${\mathcal{T}}^{3}$. This potential is a self-dual solution 
(\ref{selfduality}) of the euclidean topologically massive 
abelian gauge theory (\ref{field equation}) on ${\mathcal{T}}^{3}$. 
Thus it defines a contact structure. 

   A special contact structure on ${\mathcal{T}}^{3}$ is given 
by $b=c=0$, $a=1$ and a re-labelling: $x \leftrightarrow y$ in 
(\ref{ABCcontactgauge})

\begin{eqnarray}\label{standardT3}
A=\cos (\nu z)dx + \sin (\nu z)dy ,
\end{eqnarray}

\noindent (ignoring $\nu/g$) \cite{KY}. As defined on 
${\mathcal{S}}^{1} \times {\mathcal{R}}^{2}$ this (\ref{standardT3}),
for $\nu=1$, is associated with the space of contact elements in the 
plane ${\mathcal{R}}^{2}$ which is isomorphic to the spherized cotangent 
bundle of the plane \cite{AN}, \cite{DS, KLR}, \cite{VG}, \cite{HG1}. 

   The cases $c=a=0$ and $a=b=0$ similarly define contact structures. 
The scaling of the unmodified abc-flow is also noted in \cite{YM}.

\subsubsection{Bianchi Type $II$, $VI$ and $VII$ Spaces}

   We can also realize this (\ref{standardT3}) as a topologically 
massive gauge potential on a space with the modified left-invariant 
basis $1$-forms 

\begin{eqnarray} \label{BianchiVII0}
& & \omega^{1}=\cos (\nu z)dx + \sin (\nu z)dy ,
\hspace*{10mm} \eta_{ab}=diag(1, \, 1, \, 1) , \nonumber \\
& & \omega^{2}=\sin (\nu z)dx - \cos (\nu z)dy , \\
& & \omega^{3}=dz , \nonumber 
\end{eqnarray}

\noindent of Bianchi type $VII$. This yields the Euclidean 
metric (\ref{metric}). The Maurer-Cartan equation for the basis 
$1$-forms yields the structure constants: $C_{2 \,\, 3}^{\,\, 1}=-\nu$, 
$C_{3 \,\, 1}^{\,\, 2}=-\nu$ of the Bianchi type $VII$ group which are 
modified by the topological mass. This reduces to the group of isometries 
of the Euclidean plane for $\nu=1$. The $1$-forms $\alpha=A=\omega^{1}$ 
and $\beta=\omega^{2}$ satisfy the self-duality equation.

  Furthermore, this contact structure is related to the standard contact 
structure defined by the Darboux contact $1$-form $\alpha=dz+xdy$ via 

\begin{eqnarray}
(x, \, y, \, z) \longrightarrow \left(  z\cos (\nu y) 
+ \frac{1}{\nu} \, x\sin (\nu y), \, \,
z\sin (\nu y) - \frac{1}{\nu} \, x\cos (\nu y), \, \, y \right) ,
\end{eqnarray}

\noindent \cite{HG}. The pull-back basis $1$-forms are 

\begin{eqnarray} \label{Darboux1}
& & \omega^{1}=dz + xdy , \nonumber \\
& & \omega^{2}=\frac{1}{\nu}dx - \nu zdy , \\
& & \omega^{3}=dy , \nonumber 
\end{eqnarray}

\noindent and the metric (\ref{metric}). The euclidean, self-dual 
topologically massive Darboux gauge potential is given by 
$A=\alpha=\omega^{1}$.  

   The lorentzian version of (\ref{BianchiVII0}) is given by the modified 
left-invariant basis $1$-forms

\begin{eqnarray} \label{lorentzian}
& & \omega^{0}=\cosh (\nu z)dx + \sinh (\nu z)dy ,
\hspace*{10mm} \eta_{ab}=diag(-1, \, 1, \, 1) , \nonumber \\
& & \omega^{1}=\sinh (\nu z)dx + \cosh (\nu z)dy , \\
& & \omega^{2}=dz , \nonumber 
\end{eqnarray}

\noindent of Bianchi type $VI$ which yields the Minkowski
metric (\ref{metric}). The Maurer-Cartan equation 
yields the structure constants: $C_{1 \,\, 2}^{\,\, 0}=\nu$, 
$C_{2 \,\, 0}^{\,\, 1}=-\nu$. This reduces to the group of isometries 
of the Minkowski plane for $\nu=1$. In this case both $\alpha=A=\omega^{0}$ 
and $\beta=\omega^{1}$ satisfy the anti-self-duality equation. 

  The contact transformation to the Darboux form $\alpha=dz+xdy$ 
is given by

\begin{eqnarray}
(x, \, y, \, z) \longrightarrow \left(  z\cosh (\nu y) 
+ \frac{1}{\nu} \, x\sinh (\nu y), \, \,
-z\sinh (\nu y) - \frac{1}{\nu} \, x\cosh (\nu y), \, \, y \right) .
\end{eqnarray}

\noindent This yields the basis $1$-forms

\begin{eqnarray} \label{Darboux2}
& & \omega^{0}=dz + xdy , \nonumber \\
& & \omega^{1}=- \frac{1}{\nu}dx - \nu zdy , \\
& & \omega^{2}=dy , \nonumber 
\end{eqnarray}

\noindent with the metric (\ref{metric}). In this case the 
lorentzian, anti-self-dual topologically massive Darboux 
gauge potential is given by $A=\alpha=\omega^{0}$. 

    We can also realize the Darboux contact form $\alpha=A=dz+xdy$ as 
an euclidean, self-dual topologically massive gauge potential on a space 
with the modified left-invariant basis $1$-forms

\begin{eqnarray} \label{Darboux3}
& & \omega^{1}=dz + xdy , \hspace*{10mm} 
\eta_{ab}=diag(1, \, 1, \, 1) , \nonumber \\
& & \omega^{2}=\frac{1}{\nu}dx , \\
& & \omega^{3}=dy , \nonumber 
\end{eqnarray}

\noindent of Bianchi type $II$ and the metric (\ref{metric}). 
The Maurer-Cartan equation yields the structure constant: 
$C_{2 \,\, 3}^{\,\, 1}=-\nu$. The Bianchi type $II$ group, 
$\nu=1$, is also known as the Heisenberg group.

   A hyperbolic version of the abc-potential (\ref{ABCcontactgauge}) 
is possible only if $b=0$ with a twist of sign in $c$: 

\begin{eqnarray} \label{ABChyperbolic}
A= \left[ a \sinh (\nu z) + c \cosh (\nu y) \right] dx
+ a \cosh (\nu z) dy - c \sinh (\nu y) dz ,
\end{eqnarray}

\noindent on the Minkowski space (\ref{lorentzian}). The $1$-form $A$ 
is anti-self-dual. The case $c=0$, $a=1$ can be realized on Bianchi 
type $VI$ space (\ref{lorentzian}) as in the euclidean case. 

\subsubsection{Bianchi Type $V$ and Specialized Type $VI$, $VII$ Spaces}

   As we have already remarked, a real-valued solution on a lorentzian 
manifold or a complex-valued solution on an euclidean manifold do not 
necessarily lead to contact structures. In the three examples below 
the solution $1$-forms $\alpha$ yield $\alpha \wedge d\alpha =0$.

   If we adapt a specialized form 

\begin{eqnarray} \label{SpecialBianchiVI}
& & \omega^{0}=e^{\pm \frac{\nu}{2} \, z} 
\left[ \cosh \left( \frac{\nu}{2} \, z \right) dx + 
\sinh \left( \frac{\nu}{2} \, z \right) dy \right] ,
\hspace*{10mm} \eta_{ab}=diag(-1, \, 1, \, 1) , \nonumber \\
& & \omega^{1}=e^{\pm \frac{\nu}{2} \, z} 
\left[ \sinh \left( \frac{\nu}{2} \, z \right) dx + 
\cosh \left( \frac{\nu}{2} z \right) dy \right] ,  \\
& & \omega^{2}=dz , \nonumber 
\end{eqnarray}
  
\noindent of the modified left-invariant basis $1$-forms of Bianchi type 
$VI$ (\ref{lorentzian}) \cite{NB} with the metric (\ref{metric}) then 
$\alpha=A=\omega^{0}\pm\omega^{1}$ is an anti-self-dual solution. The 
Maurer-Cartan equation yields the structure constants: 
$C_{1 \,\, 2}^{\,\, 0}=\nu/2$, $C_{2 \,\, 0}^{\,\, 1}=-\nu/2$, 
$C_{0 \,\, 2}^{\,\, 0}=\pm\nu/2$, $C_{1 \,\, 2}^{\,\, 1}=\pm\nu/2$. 
An anisotropic version of the frame (\ref{SpecialBianchiVI}) with 
$\nu=2$ yields a solution of the topologically massive gravity \cite{NB}. 

  For the euclidean version of (\ref{SpecialBianchiVI}) with the 
specialized form 

\begin{eqnarray} \label{SpecialBianchiVII}
& & \omega^{1}=e^{\pm \frac{\nu}{2} \, z} 
\left[ \cos \left( \frac{\nu}{2} \, z \right) dx + 
\sin \left( \frac{\nu}{2} \, z \right) dy \right] ,
\hspace*{10mm} \eta_{ab}=diag(1, \, 1, \, 1) , \nonumber \\
& & \omega^{2}=e^{\pm \frac{\nu}{2} \, z} 
\left[ \sin \left( \frac{\nu}{2} \, z \right) dx -
\cos \left( \frac{\nu}{2} z \right) dy \right] ,  \\
& & \omega^{3}=dz , \nonumber 
\end{eqnarray}
  
\noindent of the modified left-invariant basis $1$-forms of Bianchi type 
$VII$ (\ref{BianchiVII0}) and the metric (\ref{metric}), the complex 
$1$-form $\alpha=A=\omega^{1}\pm i\omega^{2}$ is a self-dual solution
with a complex-valued topological mass: $\nu \longrightarrow (1-i)\nu/2$
in (\ref{selfduality}). The Maurer-Cartan equation yields the structure 
constants: $C_{2 \,\, 3}^{\,\, 1}=-\nu/2$, $C_{3 \,\, 1}^{\,\, 2}=-\nu/2$, 
$C_{1 \,\, 3}^{\,\, 1}=\pm\nu/2$, $C_{2 \,\, 3}^{\,\, 2}=\pm\nu/2$. 

   If we adapt the modified left-invariant basis $1$-forms  

\begin{eqnarray} \label{SpecialBianchiV}
& & \omega^{1}=e^{-\nu z} dx ,
\hspace*{10mm} \eta_{ab}=diag(1, \, 1, \, 1) , \nonumber \\
& & \omega^{2}=e^{-\nu z} dy ,  \\
& & \omega^{3}=dz , \nonumber 
\end{eqnarray}

\noindent of Bianchi type $V$ with the metric (\ref{metric}), then 
$\alpha=A=\omega^{0}\pm i\omega^{1}$ is a (anti-)self-dual solution
with a complex-valued topological mass: $\nu \longrightarrow \pm i\nu$
in (\ref{selfduality}). The Maurer-Cartan equation yields the structure 
constants: $C_{1 \,\, 3}^{\,\, 1}=-\nu$, $C_{2 \,\, 3}^{\,\, 2}=-\nu$. 

\subsection{The Anti-de Sitter Space} 

   An anti-self-dual, lorentzian solution of the topologically 
massive abelian gauge theory on AdS space ${\mathcal{H}}^{3}$ is 
discussed in \cite{ANS}, \cite{S1}. The AdS space is a Bianchi type 
$VIII$ space \cite{ANS}. It is globally given as 
${\mathcal{S}}^{1} \times {\mathcal{H}}^{2}_{+}$ where 
${\mathcal{H}}^{2}_{+}$ is the upper portion of a hyperboloid 
of two sheets in ${\mathcal{R}}^{3}$. The contact gauge potential
is given by

\begin{eqnarray}\label{lorentzgaugepotential}
A=-\frac{1}{2}\frac{\nu}{g} \, \omega^{3} =-\frac{1}{2}\frac{\nu}{g} 
\, \left[ d\psi + \cosh ( \nu\theta ) d\phi \right] .
\end{eqnarray}

\noindent Here

\begin{eqnarray} 
& & \omega^{1} = -\cos(\nu\psi)d\theta-\sin(\nu\psi)\sinh(\nu\theta)d\phi ,
\hspace*{10mm} \eta_{ab}=diag(-1, \, -1, \, 1) , \nonumber \\
& & \omega^{2} = -\sin(\nu\psi)d\theta+\cos(\nu\psi)\sinh(\nu\theta)d\phi , \\
& & \omega^{3} = d\psi+\cosh(\nu\theta)d\phi ,\nonumber 
\end{eqnarray}

\noindent are the modified left-invariant basis $1$-forms of $SU(1, \, 1)$ 
which is parameterized in terms of the \textit{Eulerian} (half) arclengths 
$\theta$, $\phi$ and $\psi$ corresponding to the Euler parameters 
$\tilde{\theta}=\nu\theta$, $\tilde{\phi}=\nu\phi$ and 
$\tilde{\psi}=\nu\psi$ on the AdS space ${\mathcal{H}}^{3}$ of 
\textit{radius} $r=2/\nu$, $\nu=ng^{2}$ \cite{S1}. The Maurer-Cartan 
equation for these basis $1$-forms yields the structure constants:
$C_{2 \,\, 3}^{\,\, 1}=-\nu$, $C_{3 \,\, 1}^{\,\, 2}=-\nu$, 
$C_{1 \,\, 2}^{\,\, 3}=\nu$ of the Bianchi type $VIII$ group which 
are modified by the topological mass. This is the lorentzian analogue 
\cite{Gilmore} of the euclidean solution (\ref{euclidgaugepotential}) 
on ${\mathcal{S}}^{3}$ \cite{S1}. The gauge potential $\alpha=A$ 
(\ref{lorentzgaugepotential}) defines an analogous \cite{V} lorentzian 
K-contact structure on  ${\mathcal{H}}^{3}$ \cite{GH, KD, Gu2}. The contact 
planes are time-like \cite{KD} with the conventions of \cite{S1}. Note that 
$\omega^{1}$ and $\omega^{2}$ are also anti-self-dual Beltrami (Trkalian) 
$1$-forms.

   We can also realize this using another set of modified invariant 
basis $1$-forms 

\begin{eqnarray} 
& & \omega^{1}= \frac{1}{\nu} \, \frac{1}{y} \,\,
\left[ \cos (\nu z)dx + \sin (\nu z)dy \right] ,
\hspace*{10mm} \eta_{ab}=diag(-1, \, -1, \, 1) , \nonumber \\
& & \omega^{2}=\frac{1}{\nu} \, \frac{1}{y} \,\,
\left[ \sin (\nu z)dx - \cos (\nu z)dy \right] , \\
& & \omega^{3}=\frac{1}{\nu} \, \frac{1}{y} 
\left( dx+\nu ydz \right) , \nonumber 
\end{eqnarray}

\noindent of Bianchi type $VIII$ \cite{TKH}. The Maurer-Cartan equation 
for these basis $1$-forms yields the same set of structure constants. 
The $1$-forms $\alpha=\omega^{3}$ and $\beta= \omega^{1}$, 
$\gamma=\omega^{2}$ are anti-self-dual. 

   The Bianchi type $II$, $VI$, $VII$, $VIII$ and $IX$ spaces above 
with their respective invariant contact structures determined by $\alpha=A$ 
and the metrics adapted so as to satisfy the (anti-)self-duality equation
are among the basic examples of homogeneous contact manifolds \cite{P}.

\section{Duality and Holomorphic Functions on ${\mathcal{R}}^{3}$}

   In this section, we adapt the classification of the complex Trkalian fields 
($\nu=1$) which is based on Clebsch decomposition of a $1$-form in terms of 
Monge potentials \cite{BT}. See the appendix for a brief explanation. We refer
the reader to \cite{BT} and \cite{RJ}, \cite{DJP} for a detailed discussion of 
the Clebsch decomposition. We shall focus on the cartesian case including 
the topological mass. In cartesian coordinates a real (anti-)self-dual 
$1$-form $\alpha$ can be written as the real part of the complex 
(anti-)self-dual $1$-form 

\begin{eqnarray}\label{complex1form}
a=e^{i\nu z}df ,
\end{eqnarray}

\noindent $\alpha=Re\{ a \}$. Here $z$ is the third component of the 
cartesian coordinates and $f:R^{2} \longrightarrow C$ is a complex-valued 
function on the $xy$-plane. Then the (anti-)self-duality equation 

\begin{eqnarray}\label{holomorphicity}
& & *da-\nu a=0: \,\, f_{w^{*}}=0, \hspace*{10mm} w=x+iy ,
\nonumber \\ 
\\
& & (*da+\nu a=0: \,\, f_{w}=0) , 
\nonumber 
\end{eqnarray}

\noindent yields that $f=u+iv$ is a (anti-)holomorphic function. Thus 
complex (anti-)self-dual solutions (\ref{complex1form}) of the euclidean 
topologically massive abelian gauge theory on ${\mathcal{R}}^{3}$ in 
cartesian coordinates are given by (anti-)holomorphic functions. Their 
real parts accordingly yield real contact structures. Note that we have 
a freedom of permutation of the $x$, $y$ and the $z$ coordinates for 
writing such solutions. The solution (\ref{complex1form}) has planar 
fronts along the $z$-axis as equiphase surfaces. In the Minkowski 
space with signature $(+, \, +, \, -)$ the (anti-)holomorphicity conditions 
of $f$ for the solution (\ref{complex1form}) of the (anti-)self-duality 
equation (\ref{holomorphicity}) interchange.

  For example the anti-holomorphic function $f=x-iy$ yields the contact 
$1$-form (\ref{standardT3}) 

\begin{eqnarray}\label{example1}
A=Re\{e^{i\nu z}df\}, \hspace*{10mm} f=x-iy .
\end{eqnarray}

\noindent Meanwhile for the holomorphic function  $f=x+iy$ we have 

\begin{eqnarray}\label{example2}
& & A=Re\{e^{i\nu z}df\}, \hspace*{10mm} f=x+iy, \\
& & \hspace*{4mm} = \cos (\nu z)dx - \sin (\nu z)dy . \nonumber 
\end{eqnarray}

\noindent Note that a change in the $z$-direction: $a=exp(-i\nu z)df$ 
interchanges the holomorphicity conditions since this corresponds 
to a complex conjugation in the $xy$-plane. 

   The contact structure defined by a real, self-dual $1$-form $\alpha$ is 
positively oriented: $\alpha \wedge d\alpha > 0$ (assume $\nu > 0$) while 
a contact structure defined by an anti-self-dual, real $1$-form is negatively 
oriented: $\alpha \wedge d\alpha < 0$. These are related by a change of 
orientation. Further, a real, self-dual or anti-self-dual $1$-form 
$\alpha$ respectively represents a contact structure with positive or 
negative helicity. The helicity density of the complex solution 
(\ref{complex1form}) vanishes: ${\mathcal{H}} (a)=0$. The helicity 
densities ${\mathcal{H}} (\alpha)$ and ${\mathcal{H}} (\beta)$ of 
its real and complex parts: $\alpha=Re\{a\}$, $\beta=Co\{a\}$ are 
given as

\begin{eqnarray}
{\mathcal{H}} (\alpha)={\mathcal{H}} (\beta)
=\left\{ \begin{array}{ll}
\nu(u_{x}^{2} +u_{y}^{2})*1 & \textrm{self-dual}\\
& \\
-\nu(u_{x}^{2} +u_{y}^{2})*1 & \textrm{anti-self-dual}
\end{array} \right.
\end{eqnarray}

\noindent where $*1=dx\wedge dy\wedge dz$.

\subsection{Orthogonal Contact Structures and Duality}

  We can construct real orthogonal contact structures as follows. Consider 
the complex $1$-forms  $a=e^{i\nu z}df$ and  $b=e^{-i\nu z}dg$. There exist
orthogonal contact structures in case $f=u+iv$ is holomorphic and $g=p+iq$ 
is anti-holomorphic or vice-versa. That is $a$ and $b$ are both self-dual or
anti-self-dual. In both cases the orthogonality condition reduces to a single 
equation: $\overrightarrow{\nabla}u\cdot\overrightarrow{\nabla}p
=u_{x}p_{x}+u_{y}p_{y}=0$. We can use the simplest solution: 
$b=ia^{*}$ of this equation for constructing the orthogonal
contact structures. Their real parts 

\begin{eqnarray}
& & \alpha= Re\{ a \}=\cos (\nu z)du-\sin (\nu z)dv \nonumber \\
& & \hspace*{4mm} =\left[ u_{x}\cos (\nu z)-v_{x}\sin (\nu z) \right] dx
+ \left[ u_{y}\cos (\nu z)-v_{y}\sin (\nu z) \right] dy , \nonumber \\
\\
& & \beta= Re\{ b \}=\sin (\nu z)du+\cos (\nu z)dv \nonumber \\
& & \hspace*{4mm} =\left[ v_{x}\cos (\nu z)+u_{x}\sin (\nu z) \right] dx
+ \left[ v_{y}\cos (\nu z)+u_{y}\sin (\nu z) \right] dy , \nonumber 
\end{eqnarray}

\noindent yield the orthogonal contact structures

\begin{eqnarray}
& & \xi_{\alpha}=Span\{\overrightarrow{U}, \, \overrightarrow{S} \},
\hspace*{20mm} 
\xi_{\beta}=Span\{\overrightarrow{V}, \, \overrightarrow{S} \} , 
\nonumber \\
\\
& & \overrightarrow{U}= \left( u_{y}\cos (\nu z)-v_{y}\sin (\nu z), \,\, 
-u_{x}\cos (\nu z)+v_{x}\sin (\nu z), \,\, 0 \right) , \nonumber \\
& & \overrightarrow{V}= \left( v_{y}\cos (\nu z)+u_{y}\sin (\nu z), \,\, 
-v_{x}\cos (\nu z)-u_{x}\sin (\nu z), \,\, 0 \right) , \nonumber \\
& & \overrightarrow{S}= \left( 0, \,\, 0, \,\, 1 \right) . \nonumber
\end{eqnarray}

\noindent The $1$-forms $\alpha$ and $\beta$ are related to 
$du$ and $dv$ in $uv$-plane by a $U(1)$ rotation

\begin{eqnarray} \label{SU(1,1)generators}
\left( \begin{array}{cc}
\alpha \\
\beta 
\end{array} \right)
= \left( \begin{array}{cc}
\cos(\nu z) & -\sin(\nu z) \\
\sin(\nu z) &  \cos(\nu z) 
\end{array} \right)
\left( \begin{array}{cc}
du \\
dv 
\end{array} \right) .
\end{eqnarray}

   However there exists no orthogonal structures arising from the 
distinct duality classes. Thus the orthogonality relation of 
the contact structures determined by the real parts of the set of 
complex $1$-form solutions $a=e^{\mp i\nu z}df$ separates this set 
into self-dual and anti-self-dual classes. Therefore we call these 
contact structures (anti-)self-dual. Note that a rotation of $\xi_{\beta}$
about the vector $\overrightarrow{V}$ produces another set of planes which are 
also orthogonal to $\xi_{\alpha}$. But its intersection with the solution set
only contains $\xi_{\beta}$. 

   A simple choice for the orthogonality condition is: $p_{x}=\mp u_{x}$, 
$p_{y}=\pm u_{y}$. This leads to the equations

\begin{eqnarray}
& &  \left( u_{x} \right) ^{2} - \left( u_{y} \right) ^{2} =0, \hspace*{10mm}
\left( p_{x} \right) ^{2} - \left( p_{y} \right) ^{2} =0 .
\end{eqnarray}

\noindent Then a consistency check yields the solutions 

\begin{eqnarray}\label{antiselfdual}
& & a_{1}=e^{i\nu z}d\left[ x+y-i(x-y) \right], \hspace*{10mm} 
b_{1}=e^{-i\nu z}d\left[ x-y-i(x+y) \right] , \nonumber \\
\\
& & a_{2}=e^{i\nu z}d\left[ x+y+i(x-y) \right], \hspace*{10mm} 
b_{2}=e^{-i\nu z}d\left[ x-y+i(x+y) \right] , \nonumber 
\end{eqnarray}

\noindent which are respectively self-dual and anti-self-dual. Their 
complex conjugates are also solutions. 

\section{The Curl Transformation}

  In this section, we apply the curl transformation \cite{M1} which 
is readily developed in the context of force-free magnetic fields or 
Beltrami (Trkalian) fields \cite{Ml1, Ml3} to the euclidean topologically 
massive abelian gauge theory in the vector form. We also discuss 
quantization of the topological mass: $\nu=ng^{2}$ on an example. 

  The curl transformation is based on decomposing a vector field into 
(complex) helical eigen-functions
 
\begin{eqnarray}
\overrightarrow{\chi}_{\lambda}(\overrightarrow{x}|\overrightarrow{k}) 
=\frac{1}{(2\pi)^{3/2}} e^{i \overrightarrow{k} \cdot \overrightarrow{x}}
\overrightarrow{Q}_{\lambda}(\overrightarrow{k}) ,
\end{eqnarray}

\noindent of the curl operator (\ref{Beltramivector}) 
with planar fronts along a direction determined by the vector 
$\overrightarrow{k}=(k_{1}, \, k_{2}, \, k_{3})$  \cite{R2}.  
The vectors $\overrightarrow{Q}_{\lambda}(\overrightarrow{k})$ 
are given as

\begin{eqnarray}
& & \overrightarrow{Q}_{\lambda}(\overrightarrow{k}) =-\frac{\lambda}{\sqrt{2}}
\left( \frac{k_{1}(k_{1}+i\lambda k_{2})}{k(k+k_{3})}-1, \,\,\,
\frac{k_{2}(k_{1}+i\lambda k_{2})}{k(k+k_{3})}-i\lambda, \,\,\,
\frac{k_{1}+i\lambda k_{2}}{k} \right), \hspace*{5mm} \lambda=\pm 1, 
\nonumber \\
& & \overrightarrow{Q}_{0}(\overrightarrow{k})=-\frac{\overrightarrow{k}}{k}, 
\hspace*{5mm}
k=|\overrightarrow{k}| .
\end{eqnarray}

\noindent The parameter $\lambda$ corresponds to sense of helicity states. 
These form a complex basis in Fourier space \cite{M1}. The basis vectors 
$\overrightarrow{Q}_{\lambda}(\overrightarrow{k})$ are undefined when 
$k+k_{3}=0$. The eigen-functions
$\overrightarrow{\chi}_{\lambda}(\overrightarrow{x}|\overrightarrow{k})$
of the curl operator 

\begin{eqnarray}
& & \overrightarrow{\nabla} \times 
\overrightarrow{\chi}_{\lambda}(\overrightarrow{x}|\overrightarrow{k})
=k\lambda
\overrightarrow{\chi}_{\lambda}(\overrightarrow{x}|\overrightarrow{k}), 
\nonumber \\
\nonumber \\
& & \overrightarrow{\nabla} \cdot 
\overrightarrow{\chi}_{\lambda}(\overrightarrow{x}|\overrightarrow{k})=0,
\hspace*{5mm} \lambda=\pm 1, \\
\nonumber \\
& & \overrightarrow{\nabla} \cdot 
\overrightarrow{\chi}_{0}(\overrightarrow{x}|\overrightarrow{k})
=-\frac{1}{(2\pi)^{3/2}} \,\, ik e^{i \overrightarrow{k} 
\cdot \overrightarrow{x}},
\nonumber 
\end{eqnarray}

\noindent form an orthogonal and complete set \cite{M1}. See the appendix 
for a brief account. Any vector field can be represented in terms of these 
eigen-functions in the fashion of a (vector) Fourier transform refining the 
Helmholtz decomposition \cite{M1}. According to Helmholtz theorem we can 
decompose a vector field into divergence-free 
[$\overrightarrow{\chi}_{\lambda}(\overrightarrow{x}|\overrightarrow{k})$,
$\lambda=\pm 1$] and curl-free
[$\overrightarrow{\chi}_{0}(\overrightarrow{x}|\overrightarrow{k})$]
components. The divergence-free component further consists of two 
helicity states: positive ($\lambda=+1$) and negative ($\lambda=-1$). 
The curl transformation is based on a helicity decomposition in the 
basis determined by $\overrightarrow{Q}_{\lambda}(\overrightarrow{k})$
\cite{MD1}. We refer the reader to \cite{M1} for the motivation of 
introduction of this basis. We shall use the divergence-free components 
for expressing the field and the potential and the curl-free component 
for the gauge transformation.

\subsection{The Gauge Field}

  An euclidean topologically massive field $\overrightarrow{F}$

\begin{eqnarray} \label{fieldequation}
\overrightarrow{\nabla} \times \overrightarrow{F}
-\nu \overrightarrow{F}=0 ,
\end{eqnarray}

\noindent can be expressed as 

\begin{eqnarray} \label{expansion}
\overrightarrow{F}(\overrightarrow{x})
=\sum^{\hspace*{7mm} \prime}_{\lambda} 
\overrightarrow{F}_{\lambda}(\overrightarrow{x}) ,
\end{eqnarray}

\noindent where

\begin{eqnarray} \label{helicitycomponent}
\overrightarrow{F}_{\lambda}(\overrightarrow{x})
=\frac{1}{g}
\int \overrightarrow{\chi}_{\lambda}(\overrightarrow{x}|\overrightarrow{k})
f_{\lambda}(\overrightarrow{k}) d^{3}k,
\end{eqnarray}

\noindent excluding the divergenceful component \cite{Ml1, Ml3}. 
The factor $1/g$ leads to the correct strength for the gauge potential
as we shall discuss on an example. In this decomposition each helicity 
component is given in terms of a scalar function 

\begin{eqnarray} \label{inverse}
f_{\lambda}(\overrightarrow{k})
=g\int 
\overrightarrow{\chi}^{*}_{\lambda}(\overrightarrow{x}|\overrightarrow{k})
\cdot \overrightarrow{F}(\overrightarrow{x}) d^{3}x ,
\end{eqnarray}

\noindent \cite{M1}. If we replace the expression 
(\ref{expansion}) in the equation (\ref{fieldequation}) 
we find 

\begin{eqnarray} \label{distribution}
f_{\lambda}(\overrightarrow{k})=\frac{\delta (k-\lambda\nu)}{k^{2}}
s_{\lambda}(\overrightarrow{k}),
\end{eqnarray}

\noindent using a radial delta function \cite{Ml1}. Thus an arbitrary 
solution is given entirely in terms of its transform on the sphere of 
radius $k=\lambda\nu=|\nu|$. Furthermore, only the eigen-functions for 
which $\lambda=sgn(\nu)$ contribute to the field (\ref{expansion}),
\cite{Ml1}. Then the expansion (\ref{expansion}) or (\ref{helicitycomponent})
simplifies into

\begin{eqnarray} \label{simplifiedexpansion}
& & \overrightarrow{F}_{\lambda}(\overrightarrow{x})
=\frac{1}{g}\int \overrightarrow{\chi}_{\lambda}
(\overrightarrow{x}|\lambda\nu \overrightarrow{\kappa})
s_{\lambda}(\lambda\nu \overrightarrow{\kappa}) d\Omega \\
& & \hspace*{15mm}
=\frac{1}{(2\pi)^{3/2}} \,\,  \frac{1}{g}
\int e^{i \lambda \nu \overrightarrow{\kappa} \cdot \overrightarrow{x}} 
\overrightarrow{Q}_{\lambda} (\overrightarrow{\kappa})
s_{\lambda}(\lambda\nu\overrightarrow{\kappa}) d\Omega, \nonumber
\end{eqnarray}

\noindent where $d\Omega$ is the spherical area element and 
$\overrightarrow{\kappa}=\frac{\overrightarrow{k}}{k}$ is a unit 
vector in transform space. Therefore a solution can be defined entirely 
by the value of its curl transform on the unit sphere in transform space
\cite{Ml1}. We call $s_{\lambda}$ the spherical curl transform in 
order to distinguish it from the full curl transform  $f_{\lambda}$ 
\cite{Ml3}. 

   A simple example \cite{Ml1} is given by 
$s_{\lambda}(\lambda\nu\overrightarrow{\kappa})=s_{0}\lambda h(\nu)
\delta(\overrightarrow{\kappa}-\overrightarrow{\kappa}_{0})$
where $s_{0}=\sqrt{2} (2\pi)^{3/2}$ and 
$\overrightarrow{\kappa}_{0}=(0, \, 0, \, 1)$. Here $h(\nu)$ is 
a function which will be determined in accordance with the strength 
of the gauge potential $\overrightarrow{A}_{\lambda}$. This yields 

\begin{eqnarray} \label{example3}
& & \overrightarrow{F}_{\lambda}(\overrightarrow{x})
=\frac{1}{g} \,\, h \,\, e^{i\lambda\nu z} \, (1, \, i\lambda, \, 0). 
\end{eqnarray}

  In order to exhibit the distributional character of both sides in 
(\ref{inverse}) if we multiply by $e^{ikp}$ and integrate over $k$ 
[replacing (\ref{distribution})], we find  

\begin{eqnarray} \label{sphericalcurltransform}
s_{\lambda}(\lambda\nu\overrightarrow{\kappa})
=\frac{1}{(2\pi)^{1/2}} g\nu^{2} e^{-i\lambda\nu p} 
\overrightarrow{Q}^{*}_{\lambda} (\overrightarrow{\kappa})
\cdot \overrightarrow{F}^{R}_{\lambda}(p, \, \overrightarrow{\kappa}) ,
\end{eqnarray}

\noindent where 

\begin{eqnarray} \label{radon}
\overrightarrow{F}^{R}_{\lambda}(p, \, \overrightarrow{\kappa})
=\int \overrightarrow{F}_{\lambda}(\overrightarrow{x})
\delta(p-\overrightarrow{\kappa} \cdot \overrightarrow{x}) d^{3}x, 
\end{eqnarray}

\noindent is the (vector) Radon transform of 
$\overrightarrow{F}_{\lambda}(\overrightarrow{x})$. This basically is 
the integral of $\overrightarrow{F}_{\lambda}(\overrightarrow{x})$ over 
the planes at a distance $p=\overrightarrow{\kappa} \cdot \overrightarrow{x}$ 
to the origin with unit normal $\overrightarrow{\kappa}$. Thus for a fixed 
$\overrightarrow{\kappa}$ this corresponds to a \textit{plane wave} that is 
a function constant on planes orthogonal to $\overrightarrow{\kappa}$ 
\cite{H1}, \cite{M}. The description of the spherical curl transform 
[and its inverse (\ref{simplifiedexpansion})] in terms of the Radon 
transform (\ref{sphericalcurltransform}) is both necessary and sufficient 
\cite{Ml3}. 

   We can easily check the spherical curl transform of the self-dual 
solutions $\overrightarrow{a}_{1}=(1 - i)e^{i\nu z}(1, \, i)$ 
(\ref{antiselfdual}) and $\overrightarrow{c}_{1}=\overrightarrow{b}_{1}^{*}
=i\overrightarrow{a}_{1}$. Their transforms are respectively
$s^{a}_{+1}(\nu\overrightarrow{\kappa})=s_{0}(1 - i)
\delta(\overrightarrow{\kappa}-\overrightarrow{\kappa}_{0})$
and $s^{c}_{+1}(\nu\overrightarrow{\kappa})
=i s^{a}_{+1}(\nu\overrightarrow{\kappa})$
(factor of $g$ ignored) where $s_{0}=\sqrt{2}/(2\pi)^{3/2}$ and 
$\delta(\overrightarrow{\kappa}-\overrightarrow{\kappa}_{0})$ is 
a delta function acting at the point $\overrightarrow{\kappa}_{0}
=\overrightarrow{e}_{3}$ on the unit sphere in transform space.
Meanwhile the transform of $\overrightarrow{b}_{1}
=-i\overrightarrow{a}_{1}^{*}$ is given as 
$s^{b}_{-1}(-\nu\overrightarrow{\kappa})
=i[s^{a}_{+1}(\nu\overrightarrow{\kappa})]^{*}$. 

   In the anti-self-dual case of the equation (\ref{fieldequation}) 
which contains an extra factor of $(-)$, we include this factor also 
in the equations (\ref{distribution}), (\ref{simplifiedexpansion}) and 
(\ref{sphericalcurltransform}). In this case only the eigen-functions 
with opposite helicity that is for which $\lambda=-sgn(\nu)$ contribute
to the field. A simple example is given by 
$s_{\lambda}(-\lambda\nu\overrightarrow{\kappa})=s_{0}(-\lambda)h(\nu)
\delta(\overrightarrow{\kappa}-\overrightarrow{\kappa}_{0})$
where $s_{0}=-\sqrt{2} (2\pi)^{3/2}$ and 
$\overrightarrow{\kappa}_{0}=(0, \, 0, \, 1)$. This yields 

\begin{eqnarray} \label{asdsolution}
& & \overrightarrow{F}_{\lambda}(\overrightarrow{x})
=\frac{1}{g} \,\, h \,\, e^{-i\lambda\nu z} \, (1, \, i\lambda, \, 0). 
\end{eqnarray}

\subsection{The Gauge Potential}

  The gauge potential for the field (\ref{helicitycomponent}) 
is given by 

\begin{eqnarray} \label{rawpotential}
\overrightarrow{A}_{\lambda}(\overrightarrow{x})
=\frac{1}{g} \,\, \lambda \int \overrightarrow{\chi}_{\lambda}
(\overrightarrow{x}|\overrightarrow{k})
f_{\lambda}(\overrightarrow{k}) \frac{1}{k} d^{3}k ,
\end{eqnarray}

\noindent \cite{M1}. If we replace the equation (\ref{distribution}) 
in this, we find the gauge potential 

\begin{eqnarray} \label{potential}
\overrightarrow{A}_{\lambda}(\overrightarrow{x})
=\frac{1}{g} \,\, \frac{1}{\nu} \int \overrightarrow{\chi}_{\lambda}
(\overrightarrow{x}|\lambda\nu \overrightarrow{\kappa})
s_{\lambda}(\lambda\nu \overrightarrow{\kappa}) d\Omega ,
\end{eqnarray}

\noindent for the field $\overrightarrow{F}_{\lambda}(\overrightarrow{x})$ 
(\ref{simplifiedexpansion}). This satisfies the self-duality equation 
$\overrightarrow{F}_{\lambda}=\nu \overrightarrow{A}_{\lambda}$ 
where $\overrightarrow{F}_{\lambda}=\overrightarrow{\nabla} 
\times \overrightarrow{A}_{\lambda}$. We find 

\begin{eqnarray}
\lambda f_{\lambda}(\overrightarrow{k}) \frac{1}{k}
=g \int \overrightarrow{\chi}^{*}_{\lambda}
(\overrightarrow{x}|\overrightarrow{k})
\cdot \overrightarrow{A}_{\lambda}(\overrightarrow{x}) d^{3}x ,
\end{eqnarray}

\noindent inverting the equation (\ref{rawpotential}).
If we multiply this by $e^{ikp}$ and integrate over $k$ 
[replacing (\ref{distribution})], we find  

\begin{eqnarray} 
s_{\lambda}(\lambda\nu\overrightarrow{\kappa})
=\frac{1}{(2\pi)^{1/2}} g\nu^{3} e^{-i\lambda\nu p} 
\overrightarrow{Q}^{*}_{\lambda} (\overrightarrow{\kappa})
\cdot \overrightarrow{A}^{R}_{\lambda}(p, \, \overrightarrow{\kappa}) ,
\end{eqnarray}

\noindent which can also be inferred from the self-duality equation 
and (\ref{sphericalcurltransform}). 

  The potential for the example (\ref{example3}) is

\begin{eqnarray} \label{examples}
& & \overrightarrow{A}_{\lambda}(\overrightarrow{x})
=\frac{1}{g} \,\, \frac{h}{\nu} \,\,
e^{i\lambda\nu z} \, (1, \, i\lambda, \, 0).
\end{eqnarray}

\noindent This corresponds to the complex solution (\ref{example2}) 
in differential forms for $\lambda=+1$ (ignoring the strength $h/g\nu$). 
Meanwhile for $\lambda=-1$ with an extra $(-)$ sign [see equation
(\ref{asdsolution})] in $\nu$ this corresponds to (\ref{example1}). 

   Note that the analysis from the equations (\ref{fieldequation}) to 
(\ref{radon}) can be directly applied  to the self-duality equation for 
the potential. For the abc-potential/dual-field (\ref{ABCcontactgauge}) 
the vector $\overrightarrow{\kappa}_{0}$ consists of three components along 
the $x$, $y$ and $z$-axis \cite{Ml1}.

\subsection{The Gauge Transformation}

   We need a  curl-free vector  $\overrightarrow{\nabla}U$ 
for a gauge transformation 

\begin{eqnarray} \label{gaugetransformation}
\overrightarrow{A}^{\prime}=\overrightarrow{A}
-\frac{1}{g}\overrightarrow{\nabla}U ,
\end{eqnarray}

\noindent of the potential. A gauge transformation should be consistent 
not only with the field equation (\ref{fieldequation}) but its integral 
$\overrightarrow{F}^{\prime}-\nu \overrightarrow{A}^{\prime}
=(1/g)\nu\overrightarrow{\nabla}U$ (self-duality equation 
with a source-like term) as well \cite{S, S1}. Thus we have 

\begin{eqnarray} \label{gauge}
\overrightarrow{\nabla}U(\overrightarrow{x})=\frac{1}{\nu}
\int \overrightarrow{\chi}_{0}(\overrightarrow{x}|\overrightarrow{k})
f_{0}(\overrightarrow{k}) d^{3}k .
\end{eqnarray}

\noindent If we ignore the factor $1/\nu$ here, then the 
dimensions of $f_{\lambda}(\overrightarrow{k})$, $\lambda=\pm 1$ 
in (\ref{helicitycomponent}), (\ref{rawpotential}) and 
$f_{0}(\overrightarrow{k})$ which would be included in the expression for 
a general vector field become inconsistent. Further, this is necessary for 
the equations (\ref{potential}) and (\ref{gaugetransformation}) 
to be consistent. This yields the gauge function

\begin{eqnarray}
U(\overrightarrow{x})
=\frac{1}{(2\pi)^{3/2}} \,\, i \, \frac{1}{\nu}
\int e^{i \overrightarrow{k} \cdot \overrightarrow{x}} 
f_{0}(\overrightarrow{k}) \frac{1}{k} d^{3}k.
\end{eqnarray}

   In connection with the examples (\ref{example3}) and 
(\ref{examples}), we can require the gauge function to take values 
in ${\mathcal{S}}^{1}$ with radius $r=1/\nu$ of ${\mathcal{S}}^{1} 
\times {\mathcal{R}}^{2} $ which is mentioned in sections 2.2 and 3.
Then $f_{0}(\overrightarrow{k})=f_{0}(k\overrightarrow{\kappa})=
f_{0} h(k) \frac{\delta(k-\lambda\nu)}{k^{2}}
\delta(\overrightarrow{\kappa}-\overrightarrow{\kappa}_{0})$ 
where $f_{0}=(2\pi)^{3/2}$ and $\overrightarrow{\kappa}_{0}=(0, \, 0, \, 1)$.
This yields 

\begin{eqnarray} \label{gaugeexample}
\overrightarrow{\nabla}U_{\lambda}(\overrightarrow{x})
=-\frac{h}{\nu} \,\, e^{i \lambda\nu z} \, (0, \, 0, \, 1) ,
\end{eqnarray}

\noindent where

\begin{eqnarray}
U_{\lambda}(\overrightarrow{x})=i\lambda \,\, \frac{h}{\nu^{2}}
\,\, e^{i \lambda\nu z} .
\end{eqnarray}

\noindent The (curl transform of) gauge transformation (\ref{gaugeexample})
acts at the same point as $\overrightarrow{A}_{\lambda}(\overrightarrow{x})$, 
(\ref{examples}) [$\overrightarrow{F}_{\lambda}(\overrightarrow{x})$, 
(\ref{example3})] on the transform sphere with this choice 
of $f_{0}(\overrightarrow{k})$. The dimensions of 
$s_{\lambda}(\lambda\nu\overrightarrow{\kappa})$ for the examples 
(\ref{example3}), (\ref{examples}) and $f_{0}(\overrightarrow{k})$ 
for (\ref{gaugeexample}) are consistent as required by the equation 
(\ref{distribution}). The introduction of the parameter $\lambda$ in 
$U$ is just for convenience in the examples.

   The field (\ref{simplifiedexpansion}), the potential 
(\ref{potential}) and the gauge transformation (\ref{gauge}) are 
respectively expressed in terms of the tangential and the normal 
vectors on the sphere in transform space. We can see this from 
the equations $\overrightarrow{\nabla} \cdot 
\overrightarrow{\chi}_{\lambda}(\overrightarrow{x}|\overrightarrow{k})
=i \overrightarrow{k} \cdot 
\overrightarrow{\chi}_{\lambda}(\overrightarrow{x}|\overrightarrow{k})
=0$, $\lambda=\pm 1$ [$\overrightarrow{k} \cdot 
\overrightarrow{Q}_{\lambda}(\overrightarrow{k}) =0$]
and 
$\overrightarrow{\nabla} \times 
\overrightarrow{\chi}_{0}(\overrightarrow{x}|\overrightarrow{k})
=i \overrightarrow{k}  \times 
\overrightarrow{\chi}_{0}(\overrightarrow{x}|\overrightarrow{k})
=0$, $\lambda=0$ [$ \overrightarrow{k}  \times 
\overrightarrow{Q}_{0}(\overrightarrow{k})=0$]. Thus we can think 
of the (vector) Fourier transform of the field and the potential 
as tangent to the sphere in the transform space \cite{MD1}, \cite{B} 
whereas the gauge transformation corresponds to a normal vector. The 
gauge potential $\overrightarrow{A}^{\prime}$ (\ref{gaugetransformation}) 
consists of both components. Note that, choice of $\overrightarrow{k}$
in (\ref{gauge}) determines the form of the gauge transformation
$\overrightarrow{\nabla}U(\overrightarrow{x})$. One could choose, 
as an example, another $\overrightarrow{k}$ in (\ref{gauge}) different 
from that in (\ref{simplifiedexpansion}) and (\ref{potential}) since 
the gauge transformation is just an arbitrary gradient vector. Then, 
this would again yield a normal vector. The precise form of a gauge 
transformation for a specific example depends on its geometric 
features.

\subsection{Discussion of the Example} 

  We can discuss the sense of quantization of the topological mass 
and the strength of the gauge potential on the examples (\ref{example3}), 
(\ref{examples}) and (\ref{gaugeexample}) following the reasoning in 
\cite{S}. If we choose $h(\nu)=\nu^{2}$ [$h(k)=k^{2}$] then the gauge 
potential (\ref{examples}) and its gauge transform 

\begin{eqnarray}
\overrightarrow{A}^{\prime}_{\lambda}
= \frac{\nu}{g} \,\, e^{i\lambda\nu z} \, (1, \, i\lambda, \, 1) ,
\end{eqnarray}

\noindent have the same strength $\nu/g$. Thus the strength of the gauge
potential will be given by the gauge coupling constant $\nu/g=ng$ if 
$\nu=ng^{2}$ \cite{S, S1}. This leads us to adopt a fundamental scale of 
length $R=1/g^{2}$ (radius) or $L=2\pi/g^{2}$ (perimeter) \cite{S}. We can 
write the relation $\nu=ng^{2}$ as $\nu=2\pi n/L$. If the gauge potential 
and the gauge transformation that is the factor 

\begin{eqnarray}
e^{i\lambda\nu z}=e^{i\lambda\frac{2\pi n}{L}z} ,
\end{eqnarray}

\noindent is a single-valued function of $z$ with the fundamental 
scale $L$ then $n$ has to be an integer. The fundamental length 
scale $L$ is the least common multiple of intervals over which the 
gauge potential and the transformation are single-valued and periodic 
for any integer $n$ in addition to the fact that they have a smaller period 
$l=L/n$. We can associate the integer $n$ with the winding number of a 
map $G: {\mathcal{S}}^{1}_{L} \longrightarrow  {\mathcal{S}}^{1}_{l}$ 
which winds the circle ${\mathcal{S}}^{1}_{L}$ of perimeter $L$ about 
the circle ${\mathcal{S}}^{1}_{l}$ of perimeter $l$ with locally invariant 
arclength \cite{S}. We remark that we can use a similar reasoning for the 
abc-potential (\ref{ABCcontactgauge}). This has already been discussed in 
the context of fluid dynamics. See for example \cite{YM, McP}. 

\subsection{A Source-like Term}

   Furthermore, if we are given a solution $\overrightarrow{F}$ of 
the field equation (\ref{fieldequation}) which is expressed as in
(\ref{simplifiedexpansion}), then we can introduce a source-like term 

\begin{eqnarray} \label{fieldequationwithsource}
\overrightarrow{\nabla} \times \overrightarrow{F}^{\prime}- 
\nu \overrightarrow{F}^{\prime}=\overrightarrow{J} ,
\end{eqnarray}

\noindent choosing $\overrightarrow{F}^{\prime}=\overrightarrow{F}
-(1/g)\overrightarrow{\nabla}V$ and 
$\overrightarrow{J}=(\nu/g)\overrightarrow{\nabla}V$ with 
a reasoning similar to a gauge transformation. Then we find 
$\overrightarrow{\nabla}\cdot\overrightarrow{J}
=-\nu\overrightarrow{\nabla}\cdot\overrightarrow{F}^{\prime}
=(\nu/g)\nabla^{2}V$. We can similarly express this 
term as

\begin{eqnarray} 
\overrightarrow{J}(\overrightarrow{x})=\frac{\nu}{g}
\int \overrightarrow{\chi}_{0}(\overrightarrow{x}|\overrightarrow{k})
f_{0}(\overrightarrow{k}) d^{3}k .
\end{eqnarray}

\noindent Thus a divergenceful
[$\overrightarrow{\chi}_{0}(\overrightarrow{x}|\overrightarrow{k})$]
term in (\ref{expansion}) which was previously excluded corresponds 
to a source-like term in the field equation (\ref{fieldequation}). 
This term yields a vector normal to the sphere in the transform space.
If $V$ is a harmonic function 
then: $\overrightarrow{\nabla}\cdot\overrightarrow{J}
=-\nu\overrightarrow{\nabla}\cdot\overrightarrow{F}^{\prime}
=0$. If we choose $V=1/x$ where $x= |\overrightarrow{x}|$, this yields:
$\overrightarrow{\nabla} \cdot \overrightarrow{F}^{\prime}=(4\pi/g)
\delta(\overrightarrow{x})$.

\section{Conclusion}

   We have discussed three structures in topologically massive abelian 
gauge theory. The most distinctive feature of the topologically massive 
gauge theories is the existence of a natural scale of length which is 
determined by the inverse topological mass. The abelian gauge theory 
reduces to the study of Beltrami (Trkalian) fields on a manifold with 
an adapted metric once this is recognized. 

   Thus the topologically massive abelian gauge theory on a Riemannian 
manifold defines a contact structure. This is locally contactomorphic to 
the standard contact structure defined by the Darboux form. Therefore a 
topologically massive (anti-)self-dual gauge potential or dual-field on 
a Riemannian manifold is locally given by the Darboux form with an adapted 
metric. In other words this theory on a Riemannian manifold locally has a 
unique solution up to contactomorphisms. These contact structures locally 
look alike in Darboux coordinates. In this sense, the topologically massive 
abelian gauge theory is implicitly the study of various local (gauge 
theoretic, physical etc.) aspects with an adapted metric of these 
contactomorphisms. Nevertheless these structures can possess different 
global features.

   We have presented solutions on Bianchi type $I$, $II$, $V$, $VI$, $VII$, 
$VIII$ and $IX$ spaces. We have briefly described the contact structures
defined by the gauge potential on the flat $3$-torus (Bianchi $I$), the AdS 
space (Bianchi $VIII$) and the $3$-sphere (Bianchi $IX$). The Bianchi type 
$II$, $VI$, $VII$, $VIII$, and $IX$ spaces with their respective invariant 
contact structures and the metrics adapted so as to satisfy the 
(anti-)self-duality equation are the examples of homogeneous contact 
manifolds. We have realized the Darboux contact form as a topologically 
massive gauge potential in euclidean (type $II$ and $VII$) and lorentzian 
(type $VI$) signatures. We have also presented complex-valued or lorentzian 
solutions that do not lead to contact structures on Bianchi type $V$ and 
specialized forms of Bianchi type $VI$ and $VII$ spaces.

   Then we have discussed a family of complex solutions of the euclidean 
topologically massive abelian gauge theory on ${\mathcal{R}}^{3}$ in 
cartesian coordinates. These complex (anti-)self-dual solutions are 
determined by (anti-)holomorphic functions. Their real parts accordingly 
yield real contact structures. The orthogonality relation of these contact 
structures separates the solution set into self-dual and anti-self-dual 
classes. Thus we call the respective contact structures self-dual and 
anti-self-dual.

   We have also applied the curl transformation to the euclidean 
topologically massive abelian gauge theory on ${\mathcal{R}}^{3}$. 
An arbitrary field or potential are given in terms of a vector tangent 
to a sphere whose radius is determined by the topological mass in the 
transform space. Meanwhile a gauge transformation corresponds to a vector
normal to this sphere in this space. The spherical curl transformation is 
known to be equivalent to the Radon transformation.

  Then we have discussed the sense of quantization of the topological 
mass: $\nu=ng^{2}$ on an example. Our discussion suggests the existence 
of a fundamental length scale $L=2\pi/g^{2}=2\pi n/\nu$ in topologically 
massive gauge theories. The strength of the gauge potential is given by 
the gauge coupling constant $\nu/g=ng$. The fundamental length scale is 
the least common multiple of intervals over which the gauge potential and 
the transformation are single-valued and periodic for any integer $n$ in 
addition to the fact that they have a smaller period $l=L/n$. A similar 
reasoning has already been used for the abc-flow in fluid dynamics. In 
the topologically massive gauge theories this length scale is naturally 
determined by the gauge coupling constant $g$.

  The approaches here are mostly motivated by fluid dynamics and 
plasma physics or magneto-hydrodynamics. To the knowledge of the author, 
the connection of these structures with the topologically massive gauge 
theories has been overlooked in the literature. Moreover, we have not 
yet discussed other important solutions. The three approaches here have 
relative advantages over each other from different points of view. Yet, 
they together point out richer and more interesting structures underlying 
topologically massive theories including gravity. These can contribute 
not only to the topologically massive theories but to our basic concepts 
such as space, time and field as well. 

\section*{Acknowledgments}

  The author would like to thank H. Gumral and S. Demir for reading 
the manuscript and for some corrections therein.

\appendix
\section{Classification of Trkalian Fields}

   We reproduce the following proposition from \cite{BT} in terms 
of $1$-forms.

   Proposition. Every $1$-form $\alpha$ can be written as the real 
part of the complex $1$-form

\begin{eqnarray}
a=e^{ig} df , \nonumber
\end{eqnarray}

\noindent $\alpha=Re\{ a \}$, where $g: R^{3} \longrightarrow R$ and 
$f: R^{3} \longrightarrow C$.

   Proof. It is always possible to construct the Monge potentials $g$, $h$, 
$k$ so that a $1$-form $\alpha$ can be expressed in Clebsch form

\begin{eqnarray}
\alpha=kdg+dh . \nonumber
\end{eqnarray}

\noindent If we define

\begin{eqnarray}
f=(h+ik) e^{-ig} , \hspace*{10mm} f^{*}=(h-ik) e^{ig} , \nonumber
\end{eqnarray}

\noindent we find

\begin{eqnarray}
k=-\frac{i}{2} \, \left( f  e^{ig} -  f^{*}  e^{-ig} \right) ,
\hspace*{10mm}  
h=\frac{1}{2} \, \left( f  e^{ig} +  f^{*}  e^{-ig} \right) .
\nonumber
\end{eqnarray}

\noindent Then we can write $\alpha$ as

\begin{eqnarray}
& & \alpha=\frac{1}{2} \left( e^{ig} df + e^{-ig} df^{*} \right) 
\nonumber \\
& & \hspace*{4mm} = Re\{ a \} . \nonumber 
\end{eqnarray}

   The authors of \cite{BT} impose the constraint that $a$ is 
Beltrami (Trkalian): $*da=a$ ($\nu=1$) on the $y^{1}=g$, 
$y^{2}=f$, $y^{3}=f^{*}$ system which is a diffeomorhism of flat 
${\mathcal{R}}^{3}$ with coordinates  $x^{1}$, $x^{2}$, $x^{3}$. 
They identify precisely two classes of solutions: in the cartesian 
and the spherical coordinates. In the cartesian coordinates they 
conclude

\begin{eqnarray}
ds^{2}=dg^{2}+dwdw^{*}, \nonumber 
\end{eqnarray}

\noindent where  $g=z$ and $f$ is a holomorphic function of $w=x+iy$.
We consider the cartesian case including the topological mass.

\section{Curl Transformation}

   The vectors $\overrightarrow{Q}_{\lambda}(\overrightarrow{k})$
which form a complex basis in Fourier space were introduced in 
\cite{M1}. These possess the following properties 

\begin{eqnarray}
& & \overrightarrow{Q}_{\lambda}^{*}(\overrightarrow{k})
\cdot \overrightarrow{Q}_{\mu}(\overrightarrow{k}) 
=\delta_{\lambda\mu}, \hspace*{17mm} \lambda, \, \mu =0, \, \pm 1,
\nonumber \\
& &\sum_{\lambda} \overrightarrow{Q}_{\lambda i}^{*}(\overrightarrow{k})
\overrightarrow{Q}_{\lambda j}(\overrightarrow{k}) 
=\delta_{ij}, \hspace*{10mm} i, \, j =1, \, 2, \, 3 . \nonumber
\end{eqnarray}

\noindent Thus the eigen-functions 
$\overrightarrow{\chi}_{\lambda}(\overrightarrow{x}|\overrightarrow{k})$
form an orthogonal and complete set 

\begin{eqnarray} \label{orthonormalcomplete}
\int\overrightarrow{\chi}_{\lambda}^{*}
(\overrightarrow{x}|\overrightarrow{k})
\cdot\overrightarrow{\chi}_{\mu}
(\overrightarrow{x}|\overrightarrow{k}^{\prime})
d^{3}x=\delta_{\lambda\mu}\delta(\overrightarrow{k}
-\overrightarrow{k}^{\prime}) , \nonumber \\
\nonumber \\
\sum_{\lambda} \int\overrightarrow{\chi}_{\lambda i}^{*}
(\overrightarrow{x}|\overrightarrow{k})
\overrightarrow{\chi}_{\lambda j}
(\overrightarrow{x}^{\prime}|\overrightarrow{k})
d^{3}k=\delta_{ij}\delta(\overrightarrow{x}
-\overrightarrow{x}^{\prime}) , \nonumber 
\end{eqnarray}

\noindent \cite{M1}. The vectors 
$\overrightarrow{Q}_{\lambda}(\overrightarrow{k})$
satisfy the following relations

\begin{eqnarray}
& & \overrightarrow{k} \times 
\overrightarrow{Q}_{\lambda}(\overrightarrow{k})
=-i \lambda k 
\overrightarrow{Q}_{\lambda}(\overrightarrow{k}), 
\hspace*{10mm} \lambda =0, \, \pm 1, \nonumber \\
& & \overrightarrow{k} \cdot
\overrightarrow{Q}_{\lambda}(\overrightarrow{k}) =0,  
\hspace*{33mm} \lambda =\pm 1,  
\nonumber \\
& & \overrightarrow{Q}_{\lambda}^{*}(\overrightarrow{k})
=-\overrightarrow{Q}_{-\lambda}(\overrightarrow{k}),
\nonumber \\
& & \overrightarrow{Q}_{\lambda}(\overrightarrow{k})
=\overrightarrow{Q}_{\lambda}(\overrightarrow{\kappa}) ,
\nonumber
\end{eqnarray}

\noindent which are useful in simplifying the expressions.

\end{document}